\documentclass[longnamesfirst,apjl]{emulateapj}
\usepackage{apjfonts,natbib,epsfig}
\tighten

\newcommand{\beq}{\begin{equation}}
\newcommand{\beqa}{\begin{eqnarray}}
\newcommand{\eeq}{\end{equation}}
\newcommand{\eeqa}{\end{eqnarray}}

\newcommand{\lsim}{\lesssim}
\newcommand{\gsim}{\gtrsim}

\newcommand{\lmk}{\left(}
\newcommand{\rmk}{\right)}

\begin{document}
\title{Searching for  primordial black hole dark matter with pulsar timing
arrays} 
\author{Naoki Seto  and Asantha Cooray}
\affil{Department of Physics and Astronomy, University of California, Irvine, CA 92617}

\begin{abstract}
We discuss the possibility of detecting the presence of 
 primordial black holes (PBHs), such as those that might account for
galactic dark matter, using modification of pulsar timing residuals when
 PBHs pass 
within $\sim $1000 AU and impart  impulse accelerations to the Earth.
With this technique,  PBHs with masses around  $M_{\rm PBH}\sim10^{25}$g ($\sim 0.1$ lunar mass) 
can be detected. Currently, the constraints on the abundance of such dark
 matter candidates 
 are weak.  A 30 year-long monitoring campaign with the proposed Square Kilometer  Array (SKA) 
can rule out a PBH fraction more than $\sim 1/10$  in the solar
 neighborhood in the form of  dark matter with mass $M_{\rm PBH}\sim 10^{25}$g. 
\end{abstract}

\keywords{pulsars: general---black hole physics---dark matter}

\section{Introduction}
Despite a long and intense observational quest, 
the physical identity of dark matter that accounts for most  of the mass in our galaxy is still unknown.  
Existing candidates for dark matter  can be found over many orders of
magnitude 
in mass from particle-physics scales to physically large astronomical
scales  
(see {\it e.g.} Bertone et al. 2005, Carr 2005 for reviews on
candidates).   
Among these, primordial black holes (PBHs) 
provide an important astrophysical candidate for dark matter.
Their abundance is now constrained relatively well for  masses 
above $M_{\rm PBH}\gsim10^{27}$g using the
galactic microlensing optical depth   (Alcock et al. 1998) and for $M_{\rm PBH}\lsim 10^{17}$g using the
Galactic gamma-ray background intensity compared to the expected
intensity of  
 Hawking radiation associated with  the black hole evaporation  (Carr \&
Sakellariadou 1999).

In the mass range
$10^{17} {\rm g} \lsim  M_{\rm PBH} \lsim 10^{20}{\rm g}$ the femto-lensing
technique  can be used to monitor the presence of PBHs,  depending on
structure of gamma-ray-bursts (Gould 1992).
Unfortunately, in the mass range
$10^{20} {\rm g} \lsim M_{\rm PBH} \lsim 10^{27}{\rm g}$ (or around $10^{-6}$ to 10 lunar masses), 
there is currently no preferred method to constrain PBH abundance 
(Carr \& Sakellariadou 1999).  
At these large masses, the flux of PBHs around the solar neighborhood is much 
smaller than the presumed flux of particle physics dark matter
candidates. Therefore,  direct detection of PBHs requires
detectors with  collecting areas or  cross-sections that span planetary
distances of order an
astronomical unit or larger.  In such a scenario, the presence of PBHs is inferred through
gravitational interaction between passing PBHs and  ``test masses'' 
 that can sense  impulse accelerations. With the typical distances expected for PBHs, these impulse
accelerations have very small 
amplitudes and can  be detected only with high-precision monitoring of
test mass locations.
This approach, however, does not involve complicated physical processes
nor detailed  astronomical models and enables
 us to study the relevant parameter space with simple physics.

Accurate monitoring of the location of a set of test
masses either from ground or space is now underway (or planned) for
gravitational wave studies with small strain amplitudes ({\it e.g.}
Cutler \& Thorne 2002, see also Nakamura et al. 1997, Ioka et al. 1998, Inoue \& Tanaka 2004 for
detecting gravitational waves from PBH binaries).
 It is  possible to detect PBH signatures with gravitational wave interferometers
(Seto \& Cooray 2004,    Adams \& Bloom2004). Given
the  time scale of the passage
and the fact that ground-based detectors are severely affected by
seismic noise below 10 Hz, 
these observations require space-based gravitational wave detectors,
such as 
 the Laser Interferometer Space Antenna (LISA) (Bender et al. 1998).   Assuming that the designed acceleration
noise level around $10^{-4}$Hz can be  extrapolated 
 down to $\sim 10^{-5}$Hz, the expected detection rate
for LISA is 0.01 events per year for a PBH mass of  $10^{17}$g.  While this detection 
rate with LISA is low, second-generation space detectors
might provide interesting limits (Seto \& Cooray 2004).

In this paper, we discuss an additional technique to detect PBHs focusing on 
pulsar timing  residuals.  With pulsar timing  observations, we can  measure the pulse-like
acceleration by a PBH over a time scale up to $\sim$10 yr (or equivalently down to 3nHz in frequency
space). Furthermore, with a Pulsar Timing Array (PTA) that monitors a large  number  
of pulsars to search for correlated signals buried in the noises of 
individual pulsars ( Hellings \& Downs 1983, Foster \& Backer 1990, Cordes et al. 2004, Kramer et al. 2004, Jenet et al. 2005),
 we can improve the overall sensitivity to detect PBH signatures. 
In this respect, we discuss the feasibility of using
proposed projects such as the Parkes Pulsar  Timing  Array (PPTA\footnote{See
http://www.atnf.csiro.au/research/pulsar/psrtime.}) and the Square
Kilometer Array (SKA\footnote{See http://www.skatelescope.org.}) to study what constraints they  place
on the fraction of PBHs that make up the dark matter.
While we limit our discussion in this paper to   dark matter candidates in
the form of PBHs, 
our calculations equally apply for other   dark matter candidates that are
sufficiently compact and have masses that are similar to the values we
study  here.

\section{Expected Detection Rate}

First, we study the passage of a PBH with mass $M_{\rm PBH}$, velocity $V$, and closest
approach distance $D$ from the Sun. A schematic diagram of the
interaction is given in Figure~1.
Hereafter,  the velocity $V$ is fixed at $V=350{\rm km~sec^{-1}}$ which is the typical velocity 
for halo dark matter relative to the solar system (Carr \& Sakellariadou
1999). Besides the  combination of parameters $(M_{\rm PBH},D)$, we can also characterize the 
event  with  the time scale and the amplitude of the gravitational perturbation induced by the PBH.

The time scale of the gravitational interaction is 
\beq
T\equiv {D}/{V} =13.6{\rm yr} \lmk {D}/{1000 {\rm AU}} \rmk
\lmk  {V}/{350 {\rm km~sec^{-1}}} \rmk^{-1} . \label{time}
\eeq
As we shall see later,  the detectable distance range $D$ from the Sun
for the upcoming pulsar timing arrays is around
1000 AU. For a circular orbit with radius $r$ around the Sun, the orbital period is
$170 (r/30{\rm AU})^{3/2}$yr and the orbital velocity is $5(r/30{\rm
AU})^{-1/2} {\rm km~sec^{-1}}$. Note that $30$ AU is the distance between Neptune and
the Sun. Therefore, both the gravitational focusing of a PBH by the Sun is
negligible  and the time scale of gravitational perturbation by a bound
object at distance $\gsim 100$AU is much longer than a realistic
expected 
observational  period which can be as long as several decades (see also Zakamska \& Tremaine 2005).

We summarize our parameters and results in Figure~2.
In this figure the time scale for interaction  $T$ is shown as the right-hand vertical
axis taking the velocity $V$ to be $350{\rm km~sec^{-1}}$. We also show two 
observational periods with $T_{\rm obs}=1$ and 10 yr using horizontal dashed lines. 
As the relevant distance $D$ from the Sun is much larger than those
between planets and the Sun, the acceleration by the PBH should be
regarded as that for the solar system barycenter (SSB).  The tidal
acceleration of the Earth relative to the SSB is suppressed by a factor
$(1{\rm AU}/D)$ and is further reduced when
integrating many orbital cycles. Therefore, we focus on detecting the
bulk acceleration of the SSB by a passing PBH using  pulsar
timing residuals (see {\it e.g.} Damour \& Taylor 1991 for a discussion on the constant SSB acceleration).   

With the time scale established, we now discuss the amplitude of the gravitational
perturbation induced by a passing PBH which, in turn, leads to a signal in
Time-of-Arrival (TOA) residuals of pulsars. The amplitude of the pulse-like acceleration
of a  passing PBH with mass 
$M_{\rm PBH}$, distance $D$, and 
velocity $V$ is $\sim GM_{\rm PBH}/D^2$, with the duration given by $\sim T=D/V$.
Thus, the TOA residual $t_r$ induced by a PBH is
\beq
t_r= \frac{\delta x}{c}\sim  \frac1c 
\frac{GM_{\rm PBH}}{D^2} T^2\sim \frac{GM_{\rm PBH}}{cV^2}, \label{amp}
\eeq
where the quantity $\delta x$ is the  modulation of the SSB's position
by the passage of a PBH. For a given velocity $V$, we have a one-to-one correspondence between
the mass of a PBH  $M_{\rm PBH}$ and the pulsar timing residual $t_r$.
Using appropriate numbers, the timing residual is
\beq
t_r=18   \lmk \frac{M_{\rm PBH}}{10^{25} {\rm g}} \rmk  \lmk  \frac{V}{350 {\rm km ~sec^{-1}}} \rmk^{-2} {\rm ns}. \label{amp2}
\eeq
In Figure 2 the residual $t_r$ is indicated by the top
horizontal axis on the
mass-distance plane, and again taking
$V=350{\rm km~sec^{-1}}$.   From the observed time  scale $T_{\rm obs}$ and  the timing residual
$t_r$, we can inversely estimate the distance $D$ and the mass $M_{\rm
PBH}$,  assuming a
magnitude for the velocity $V$. Note that, in a general situation,
 there are two observable quantities with three free parameters. It is difficult to determine all three quantities 
 $D$, $M_{\rm PBH}$ and $V$ simultaneously by observing a pulse-like acceleration, unless
one of these parameters can be established independently. Here we fix the velocity, but with any
detection, this degeneracy can impact a simple interpretation of an event.

Next we  estimate the flux of nearby PBHs. The density
$\rho_{\rm PBH}$ 
of PBH around the Sun must be smaller than the estimated local density
of dark matter $\rho_{\rm DM}=0.011\pm 0.005 [M_\odot {\rm pc^{-3}}]$
(Olling \& Merrifield 2001). We 
denote the fraction of dark matter in the form PBHs as $f_{\rm PBH}=\rho_{\rm PBH}/\rho_{\rm DM}$.
 For a  given mass $M_{\rm PBH}$, the event rate  $R$ of PBH  passing near
the Sun  within a distance $D$ is 
$
R\equiv {\pi \rho_{\rm PBH} D^2 V}/{ M_{\rm PBH}}, 
$
or 
\beqa
R&=& 0.068 {\rm yr^{-1}} \lmk \frac{f_{\rm PBH}}{1.0} \rmk \lmk
\frac{\rho_{\rm DM}}{0.011 M_\odot {\rm pc^{-3}}} \rmk 
\nonumber \\
& &\times  \lmk  \frac{V}{\rm 350 km~sec^{-1}} \rmk
\lmk\frac{M_{\rm PBH}}{\rm 10^{25} g} \rmk^{-1} \lmk\frac{D}{\rm 1000 AU} \rmk^2. \label{rate}
\eeqa
In Figure 2 we show the distance $D$ for the given event rates $R=0.1$ yr$^{-1}$ and
$1{\rm yr^{-1}}$  under the assumption that $f_{\rm PBH}=1$  ($\rho_{\rm PBH}=\rho_{\rm DM}$).
Here, note that the distance depends on the mass as $D\propto M_{\rm PBH}^{-1/2}$.

If our observational time period for the monitoring of pulsars $T_{\rm obs}$ is less than $\sim 10$ yr, we need to
search  for events below the upper short-dashed line in Figure 2. At the
same time, for an event to be present given the expected event rate, it should be above the lower long-dashed line.
The resulting mass range is $M_{\rm PBH}\lsim 4\times 10^{25}$g corresponding to a TOA residual of $t_r \lsim
7$ns. This is much smaller than the expected noise level $\sim 100$ns
for relatively stable pulsars with a 10 yr observational period ({\it e.g.} Jenet
et al. 2005).  Therefore, it is statistically difficult to detect a
passing PBH 
using a single pulsar. The detection might become possible with a 
pulsar timing array (PTA) where one can analyze a large number of pulsars to search
for a correlated timing residuals below the  noise level of individual
pulsar.


We now discuss the possibility of using PTA projects for a detection of
PBHs. 
Pulsar timing observations have been long used to search for
the low-frequency gravitational wave background. The basic theoretical
framework is well developed  (see Detweiler 1979, Hellings \& Downs 1983,
Blandford et al. 1984, Foster \& Backer 1990, Damour \& Taylor 1991, Kaspi et al. 1994, Thorsett \& Dewey 1996, McHugh
et al. 1996). 
A pulse-like gravitational wave with a time scale $T$ and a non-dimensional amplitude $h_c$
modulates  radio signals from a pulsar by $\Delta
\nu/\nu\sim h_c $ ($\nu$: frequency of the radio pulse).
The same nondimensional observable 
$s_c\equiv \Delta \nu/\nu$ generated by the PBH acceleration is given as
\beq
s_c= \frac{\delta \nu}{\nu}\sim \frac{\delta v}{c}\sim \frac1c 
\frac{GM_{\rm PBH}}{D^2} T\sim \frac{GM_{\rm PBH}}{cDV}, \label{amp3}
\eeq
where the quantity $\delta v$ is the velocity modulation of the SSB by the
PBH.  
This expression  is evaluated as  
$
s_c=4.3\times 10^{-17}  \lmk \frac{M_{\rm PBH}}{10^{25} {\rm g}} \rmk \lmk \frac{D}{1000 {\rm AU}}
\rmk^{-1}  \lmk  \frac{V}{350 {\rm km ~sec^{-1}}} \rmk^{-1} . 
$
In Figure 2, amplitude  $s_c$ is shown with solid lines.    Here,  we discuss the
modulation of pulsars' signals over a given time scale $T$. For reference, note that TOA residual $t_r$ and
the nondimensional amplitude $s_c$ are almost equivalent in  Fourier
space where we have $t_r \cdot f=s_c$ with $f\sim T^{-1}$,  as seen with 
eqs.(\ref{amp}) and (\ref{amp3}). 

For each pulsar, the TOA modulation due to the gravitational wave background is given by two terms,
the pulsar term  and the Earth term. The magnitude of  these two contributions are comparable. 
The pulsar term is determined by the gravitational wave background at
the pulsar, while the Earth term 
correlates the TOA residuals for all pulsars (Detweiler 1979, see also Jenet et al. 2003). The pulsar term can be
regarded as  noise when searching for the correlated signal by the gravitational wave background with a PTA. 
The TOA residual by a passing  PBH around the Earth is completely
correlated among pulsars.
 This is
advantageous when searching for a PBH signature, especially in the case
when  the signal is strong enough to be detected 
with a small number of  pulsars.  But this does not hold when noise of
each pulsar 
dominates the  
correlated signal and we need a  PTA to detect it.

Here we calculate the expected sensitivity $t_r$ as a function of the signal duration $T$, or
equivalently, we examine  the  region in mass-distance plane (Figure 2) that can be probed
with upcoming projects. To estimate the measurement sensitivity
$t_r(T)$ as a function of the time scale $T$,  we first
summarize results for the sensitivity $h_c(T)$ given for an observation
of the  
gravitational wave background with a PTA.  This will give us an estimation
of the  measurement sensitivity for a PBH event  in terms of $s_c(T)$. Then, we 
can derive the sensitivity in terms of $t_r(T)$ using the simple
correspondence between $s_c$ and $t_r$ mentioned before.  
In the next section, we will discuss how we can separate PBH signatures from signatures related to a
gravitational wave background.

Because a PBH signal has a single pulse-like profile with a finite
duration $T$, one might expect that the detection significance for the event
is not improved with an observational period $T_{\rm obs}$ much longer than the signal
duration $T$. In reality, parameters of pulsars such as their
directions and distances 
can be estimated better with a long observational
period  and, consequently, timing residuals are less
affected by errors in these  parameters (Blandford et al. 1984). 
While this  dependence on the
observational period  $T_{\rm obs}$ is
important for $T_{\rm obs}\lsim 3$yr, we neglect it with the assumption of a sufficient
observational period $T_{\rm obs}\ge 5$yr.

\begin{figure}[!t]
\epsscale{1.1}
\plotone{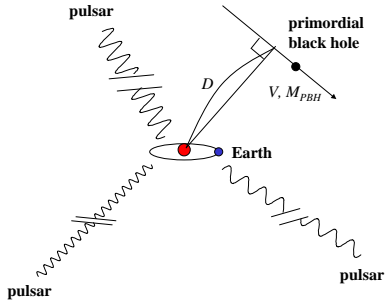}
\caption{Schematic picture for PBH search with pulsar timing observation. A PBH
 with mass $M_{\rm PBH}$ and  velocity $V$ passes near the Sun with closest
 distance $D\gg 1$AU. We fix the velocity $V$ at its typical value
 350$\rm km~sec^{-1}$. }
\label{fig1}
\end{figure}

\begin{figure}[!t]
\epsscale{1.1}
\plotone{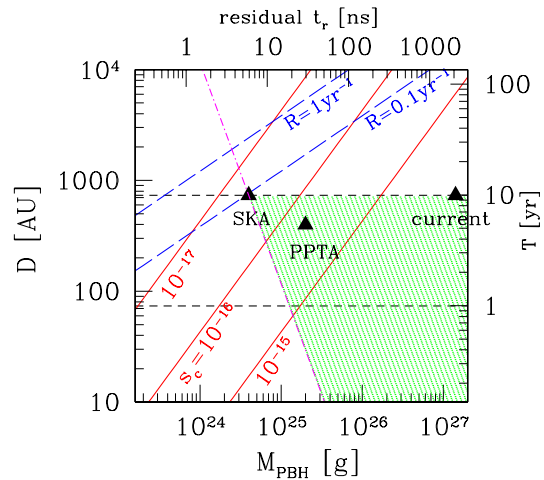}
\caption{Mass-distance relations under various observational conditions
 for PBH search. The short-dashed lines $(D\propto M_{\rm PBH}^0)$ show the signal
 duration for the fixed PBH velocity $V=350$$\rm km~sec^{-1}$. We have one to one correspondence between the mass $M_{\rm PBH}$ (bottom axis) and the residual time $t_r$ (top axis). The same is true for the distance $D$ (right axis) and the time scale $T$ (left axis).  The solid lines
 $(D\propto M_{\rm PBH}^1)$ show the amplitudes $s_c$ of timing residuals in the
 form $\Delta \nu/\nu$.  The triangles correspond to the points $(T_{\rm obs},
\Omega_{GW}(T_{\rm obs}^{-1})h^2)$ for the following three pulsar timing projects, SKA, PPTA and
 current results. The hatched region is detectable event for SKA bounded
 by the short-dashed and dash-dotted lines. The triangle for SKA moves along
 the dash-dotted line $(D\propto M_{\rm PBH}^{-2})$ for different observational
 period $T_{\rm obs}$.  The long-dashed lines are distances $D$ for given
 event rates $R$ and mass $M_{\rm PBH}$ with putting $\rho_{\rm PBH}=\rho_{\rm DM}=0.011M_\odot {\rm pc^{-3}}$. }
\label{fig2}
\end{figure}

For studies involving the gravitational wave background, we often use the spectrum
$\Omega_{GW}(f) \equiv \frac1\rho_c \frac{d\rho_{GW}(f)}{d\ln f}$ to describe
the energy density of  waves per logarithmic frequency interval
around a  frequency $f$,  when normalized by  the critical  density $\rho_c$  of the
universe (see {\it e.g.} Allen 1997, Maggiore 2000). The characteristic amplitude $h_c(T)$ is related to
$\Omega_{GW} (f)$ at $T=f^{-1}$ as
$
\Omega_{GW}(f)=\frac{2\pi^2}{3 H_0^2} f^2 h_c(T)^2,
$
($H_0=100h{\rm km~sec^{-1} Mpc^{-1}}$: the Hubble parameter) and we can write
\beq
h_c(T)=4.0\times 10^{-16}\lmk \frac{\Omega_{GW}(f=T^{-1})h^{2}}{10^{-10}} \rmk^{1/2}
\lmk\frac{T}{\rm 10 yr} \rmk.\label{omega}
\eeq

For  a given observational period $T_{\rm obs}$, it is difficult to extract
information on gravitational waves with periods longer than observational duration.
This is because we cannot separate TOA residuals due to gravitational waves from  
secular effects such as  the
long-term spin evolution of pulsars (Kaspi et al. 1994, see also Damour
\& Taylor 1991). With the pulsar timing analysis,  the
sensitivity of the spectrum $\Omega_{GW}(f)$  generally becomes best at
the frequency $f\simeq 1/T_{\rm obs}$ or time scale $T\simeq T_{\rm obs}$. 
In terms of the characteristic amplitude $h_c(T)$, the expected
sensitivity at  $T=T_{\rm obs}$ can be written with the
scaling relation  $h_c(T_{\rm obs})\propto T_{\rm obs}^{-3/2}$ (Rajagopal \& Romani
1995, Jaffe \& Backer 2003), with an overall normalization that depends
on the  parameters of the individual PTA.
Similarly, we have $s_c(T)\propto T^{-3/2}$ or
$t_r(T)\propto T^{-1/2}$ for the PBH acceleration
with  $T<T_{\rm obs}$. 
From eqs.(\ref{time}) and (\ref{amp}),  the  above scaling relation
becomes 
$D\propto M_{\rm PBH}^{-2}$ in the mass-distance plane.

For the present analysis the sensitivities of various pulsar timing projects
can be characterized by the combination $(T_{\rm obs}$,
$\Omega_{GW}(T_{\rm obs}^{-1})h^2)$  associated with a gravitational
wave background search.  
Here we use values $(T_{\rm obs},
\Omega_{GW}(T_{\rm obs}^{-1})h^2)=(5{\rm yr}, 9.1\times 10^{-11})$ for
the two parameters   for the PPTA following Jenet et al. (2006) 
($\Omega_{GW}h^2$: 95\% detection rate), and $(10{\rm yr}, 10^{-12.6})$
for the SKA-PTA from Damour
and Vilenkin (2005) ($\Omega_{GW}h^2$: 6-$\sigma$ detection
significance). As an added reference, we also take $(10{\rm yr}, 6.0\times 10^{-8})$
for the current observational sensitivity from Jenet et al. (2006). 
In Figure 2, we show the corresponding points $(T_{\rm obs},
\Omega_{GW}(T_{\rm obs}^{-1})h^2)$ as filled triangles on the mass-distance
plane using eq.(\ref{omega}). For  the SKA-PTA project, the detectable event rate is also shown as the hatched
region. This region is bounded by the two lines related to the  requirement 
$T<T_{\rm obs}$  and another related to $D\propto M_{\rm PBH}^{-2}$.

The triangle for the  SKA-PTA project with $T_{\rm obs}=10$ yr is nearly on the long-dashed line $R=0.1{\rm
yr^{-1}}$. This means that with $f_{\rm PBH}=1$ and $M_{\rm PBH}\sim 3\times 10^{24}$g,
we will be able to detect one PBH event in a 10 yr observation with a
duty cycle $C=1$.  With an observational period $T_{\rm obs}$   $\sim 30$ 
yr, the triangle for the  SKA-PTA moves upward along the dash-dotted
line and is now
nearly on the long-dashed line for $R=1{\rm yr^{-1}}$ and mass $M_{\rm PBH}\sim
2\times 10^{24}$g. This indicates that  if the local
dark matter is made by PBHs with $f_{\rm PBH}=1$ and $M_{\rm PBH}\sim 2\times 
10^{24}$g,  the pulse-like signals for passing PBHs with time scales
$T\sim T_{\rm obs}\sim30$ yr will  highly overlap in data streams
of pulsar timing.

The duty cycle $C$
is
proportional to $D^2 T/M_{\rm PBH}\propto T^3 /M_{\rm PBH} $ for  events
with  a time scale $T$ and a mass $M_{\rm PBH} (\propto t_r)$. If a point  in
the detectable zone (the hatched region in Figure 2) on the mass-distance plane is
expected to have
$C>1$ as is the case for the previous example, there is an appropriate PBH mass above which the
duty cycle is lower than 1, and events will be resolvable for the same
time scale $T$. We can also decrease the
expected  duty
cycle  by searching  shorter time-scale events and keeping the
target mass $M_{\rm PBH}$. In  this case the new point with $C<1$ on
the plane  might 
be out of the detectable zone after crossing the left boundary
(dash-dotted line)  from above.

On the other hand, if no PBH signatures are detected with the SKA-PTA in
30 yr, we can
constrain their abundance fraction down to $f_{\rm PBH} \lsim 1/10$ around $M_{\rm PBH}\sim2\times 10^{24}$g.
Of course, this is based on an  order of magnitude estimate and  we
need detailed analysis to evaluate the exact fraction. While we have
discussed the SKA-PTA so far, using the
PPTA over an observational period $T_{\rm obs}$ of 5 yr, it  will be difficult to detect a single
PBH. With $T_{\rm obs}\sim 20$ yr, the PPTA has the
potential to provide a useful constraint around $M_{\rm PBH}\sim 10^{25}$g.

\section{Discussions}

As we have discussed so far, it is quite important to have a long
observational period $T_{\rm obs}$ and analyze long-term TOA residuals when
searching for PBH signatures. For this purpose, we need stable reference clocks and high-precision
ephemerides of the solar system. While we do not discuss required accuracies of them, 
 those requirements are  similar to those that are needed for gravitational wave
searches (Foster \& Backer 1990, Kaspi et al. 1994, Hobbs et al. 2006). 

Merging super-massive black hole binaries are considered to be the dominant
astrophysical source of the gravitational wave background around a frequency of
$\sim 0.1{\rm yr^{-1}}$ (or a time scale $T\sim 10$ yr) (see also Damour \&
Vilenkin 2005 for the gravitational wave background generated by cosmic strings). The estimated
amplitude  using simple assumptions is $h_c\sim 10^{-15.5}-1
0^{-16}$ at $f=
0.1{\rm yr^{-1}}$ (Jaffe \& Backer 2003, Wyithe \& Loeb 2003). But depending on model parameters, 
a smaller amplitude $h_c\sim 10^{-16.5}$ is also predicted (Enoki \& Nagashima 2006). While the estimated 
amplitude has a large  uncertainty at present, this gravitational wave background can be an effective
noise for a PBH  search and vice-versa. Therefore, it is desirable to
observationally distinguish these two signatures in TOA residuals.

One straightforward approach is to investigate spectral information of
the residuals  expected for the PBH accelerations and at the same time  the gravitational wave background.  
Another approach  is to use the angular pattern of the
residuals on the sky.  A PBH signature will have a dipole pattern
($l=1$) whose 
direction is determined by the acceleration vector.
In this respect the dipole component parallel to the ecliptic plane will
be more affected by errors in the
solar system  ephemerides  than  the component perpendicular to the
plane.  
On the other hand, the angular pattern due to the gravitational wave
background does not have a dipole moment but has multipole modes starting
from  the quadrupole $(l=2)$ where 75\% of the angular power is expected to exist (Burke 1975).
Therefore, by studying the angular pattern of the timing residuals we
can, in principle, separate the signatures of PBH events from gravitational
waves. Once PTAs start to monitor sufficient pulsars, it is clear these topics must be further investigated.


NS would like to thank S. Kawamura, T. Tanaka and T. Daishido for
stimulating conversations, and M. Kramer for information on SKA. We also
thank  an anonymous referee for
detailed comments on the manuscript and J. Cooke for carefully
reading the draft. This work is supported by McCue fund at UC Irvine.


\begin{thebibliography}{99}

\bibitem[Adams \& Bloom(2004)]{2004astro.ph..5266A} Adams, A.~W., \& Bloom, 
J.~S.\ 2004,  arXiv:astro-ph/0405266 

\bibitem[Alcock et al.(1998)]{} Alcock, C., et al.\ 
1998, \apjl, 499, L9 

\bibitem[Allen(1997)]{1997stgr.proc....3A} Allen, B.\ 1997,  arXiv:gr-qc/9604033

\bibitem[lisa]{lisa} Bender, P. L., et al. 1998, LISA Pre-Phase A Report (2d ed.; Garching:
 Max-Plank-Institut fur Qantenoptik)

\bibitem[Bertone et al.(2005)]{2005PhR...405..279B} Bertone, G., Hooper, 
D., \& Silk, J.\ 2005, \physrep, 405, 279 




\bibitem[Blandford et al.(1984)]{1984JApA....5..369B} Blandford, R., 
Romani, R.~W., \& Narayan, R.\ 1984, Journal of Astrophysics and Astronomy, 
5, 369 


\bibitem[Burke(1975)]{1975ApJ...196..329B} Burke, W.~L.\ 1975, \apj, 196, 
329 

\bibitem[Carr \& Sakellariadou(1999)]{1999ApJ...516..195C} Carr, B.~J., \& 
Sakellariadou, M.\ 1999, \apj, 516, 195 




\bibitem[Carr(2005)]{2005astro.ph.11743C} Carr, B.~J.\ 2005, arXiv:astro-ph/0511743 





\bibitem[Cordes et al.(2004)]{2004NewAR..48.1413C} Cordes, J.~M., et al. 2004, 
New Astronomy Review, 48, 1413 

\bibitem[Cutler \& Thorne(2002)]{2002gr.qc.....4090C} Cutler, C., \& 
Thorne, K.~S.\ 2002,  arXiv:gr-qc/0204090 




\bibitem[Damour \& Taylor(1991)]{1991ApJ...366..501D} Damour, T., \& 
Taylor, J.~H.\ 1991, \apj, 366, 501 

\bibitem[Damour \& Vilenkin(2005)]{2005PhRvD..71f3510D} Damour, T., \& 
Vilenkin, A.\ 2005, \prd, 71, 063510 


\bibitem[Detweiler(1979)]{1979ApJ...234.1100D} Detweiler, S.\ 1979, \apj, 
234, 1100 



\bibitem[Enoki \& Nagashima(2006)]{2006astro.ph..9377E} Enoki, M., \& 
Nagashima, M.\ 2006,  arXiv:astro-ph/0609377 


\bibitem[Foster \& Backer(1990)]{1990ApJ...361..300F} Foster, R.~S., \& 
Backer, D.~C.\ 1990, \apj, 361, 300 




\bibitem[Gould(1992)]{1992ApJ...386L...5G} Gould, A.\ 1992, \apjl, 386, L5 




\bibitem[Hellings \& Downs(1983)]{1983ApJ...265L..39H} Hellings, R.~W., \& 
Downs, G.~S.\ 1983, \apjl, 265, L39 

\bibitem[Hobbs et al.(2006)]{2006MNRAS.369..655H} Hobbs, G.~B., Edwards, 
R.~T., \& Manchester, R.~N.\ 2006, \mnras, 369, 655 



\bibitem[Inoue \& Tanaka(2003)]{2003PhRvL..91b1101I} Inoue, K.~T., \& 
Tanaka, T.\ 2003, Phys. Rev. Lett., 91, 021101 

\bibitem[Ioka et al.(1998)]{1998PhRvD..58f3003I} Ioka, K., et al. 1998, \prd, 58, 063003 





\bibitem[Jaffe \& Backer(2003)]{2003ApJ...583..616J} Jaffe, A.~H., \& 
Backer, D.~C.\ 2003, \apj, 583, 616 


\bibitem[Jenet et al.(2004)]{2004ApJ...606..799J} Jenet, F.~A., et al. \ 2004, \apj, 606, 799 




\bibitem[Jenet et al.(2005)]{2005ApJ...625L.123J} Jenet, F.~A., et al. 2005, \apjl, 625, L123 


\bibitem[Jenet et al.(2006)]{2006astro.ph..9013J} Jenet, F.~A., et al.\ 
2006,  arXiv:astro-ph/0609013 




\bibitem[Kaspi et al.(1994)]{1994ApJ...428..713K} Kaspi, V.~M., Taylor, 
J.~H., \& Ryba, M.~F.\ 1994, \apj, 428, 713 



\bibitem[Kramer et al.(2004)]{2004NewAR..48..993K} Kramer, M., et al. 
2004, New Astronomy Review, 48, 993 





\bibitem[Maggiore(2000)]{2000PhR...331..283M} Maggiore, M.\ 2000, \physrep, 
331, 283 



\bibitem[McHugh et al.(1996)]{1996PhRvD..54.5993M} McHugh, M.~P.,  et al. 1996, \prd, 54, 5993

\bibitem[Nakamura et al.(1997)]{1997ApJ...487L.139N} Nakamura, T., et al.\ 1997, \apjl, 487, L139 

\bibitem[Olling \& Merrifield(2001)]{2001MNRAS.326..164O} Olling, R.~P., \& 
Merrifield, M.~R.\ 2001, \mnras, 326, 164 


\bibitem[Rajagopal \& Romani(1995)]{1995ApJ...446..543R} Rajagopal, M., \& 
Romani, R.~W.\ 1995, \apj, 446, 543 


 

\bibitem[Seto \& Cooray(2004)]{2004PhRvD..70f3512S} Seto, N., \& Cooray, 
A.\ 2004, \prd, 70, 063512 




\bibitem[Thorsett \& Dewey(1996)]{1996PhRvD..53.3468T} Thorsett, S.~E., \& 
Dewey, R.~J.\ 1996, \prd, 53, 3468 


\bibitem[Wyithe \& Loeb(2003)]{2003ApJ...590..691W} Wyithe, J.~S.~B., \& 
Loeb, A.\ 2003, \apj, 590, 691 




\bibitem[Zakamska \& Tremaine(2005)]{2005AJ....130.1939Z} Zakamska, N.~L., 
\& Tremaine, S.\ 2005, \aj, 130, 1939 




\end{thebibliography}
\end{document}